\documentstyle[sprocl,psfig]{article}

\input{psfig.sty}

\def\be{\begin{equation}}
\def\ee{\end{equation}}
\def\bea{\begin{eqnarray}}
\def\eea{\end{eqnarray}}
\begin{document}

\title{POLARIZED PARTON DISTRIBUTIONS IN LIGHT-FRONT DYNAMICS}

\author{P. Faccioli}

\address{ECT$\star$, Villa Tambosi, 38050 Villazzano,\\
(Trento), Italy\\E-mail: faccioli@ect.it }

\author{F. Cano}

\address{Dipartimento di Fisica, Universit\`a degli Studi di Trento, \\
 I-38050 Povo (Trento) Italy \\ E-mail: cano@science.unitn.it }  

\author{M. Traini}

\address{Dipartimento di Fisica, Universit\`a degli Studi di Trento, \\
and Istituto Nazionale di Fisica Nucleare, G.C. Trento \\
 I-38050 Povo (Trento) Italy \\ E-mail: traini@science.unitn.it }  

\author{V. Vento}

\address{Departament de F\'{\i}sica Te\`orica, Universitat 
de Val\`encia\\
and Institut de F\'{\i}sica Corpuscular, Centre Mixt 
Universitat de Val\`encia,\\
Consejo Superior de Investigaciones Cient\'{\i}ficas,\\
E-46100 Burjassot (Val\`encia) Spain\\
E-mail: vicente.vento@uv.es }


\maketitle\abstracts{ We present a consistent calculation of the
structure functions within a light-front constituent quark model of
the nucleon. Relativistic effects and the relevance of the covariance
constraints are analyzed for polarized parton distributions.  Various
models, which differ in their gluonic structure at the hadronic scale,
are investigated. The results of the full covariant calculation are
compared with those of a non-relativistic approximation to show the
structure and magnitude of the differences. It is also shown how
measurements of transversity in doubly polarized Drell-Yan lepton pair
production are a clearcut sign of covariance requirements for the
spin. }

\section{Introduction}

We report here on calculations of the polarized structure functions
within a covariant quark model of the nucleon.  The framework used is
based on a quark-parton picture we have developed to describe the
structure on the nucleons in the deep inelastic regime
\cite{noi2}. The central assumption in this procedure
\cite{PARISI76} is the existence of a scale,
$\mu_0^2$, where the short range (perturbative) part of the
interaction is negligible, therefore the glue and sea are suppressed,
and the long range (confining) part of the interaction produces a
proton composed of three (valence) quarks only. Jaffe and Ross
\cite{JAFFE80} proposed thereafter, to ascribe the quark model
calculations of matrix elements to that hadronic scale
$\mu_0^2$. For larger $Q^2$ their Wilson coefficients will give the
evolution as dictated by perturbative QCD. In this way quark models,
summarizing a great deal of hadronic properties, may substitute {\it ad hoc}
low-energy parameterizations.
 
However, it has been quite evident, since the original formulation of
many of these models, that relativistic effects to the nucleon wave
function, as well as covariance requirements, are needed even for a
phenomenological description of the structure of hadrons.  To
accomplish with this aim we develop a constituent quark model in the
light-front realization of the Hamiltonian dynamics where covariance
can be incorporated in a rather transparent and elegant manner \cite{FACCIOLI99}.

	We apply this formalism to the calculation of polarized parton
distribution. The role of initial soft gluons in the description of
the parton distribution (and especially gluon polarization) is
analyzed. Finally we show how differences between transverse and
longitudinal polarization are a direct consequence of a light-front
treatment of the spin and they may be detected experimentally in a
rather model independent way.

\section{Polarized parton distributions at the hadronic scale}

The parton distributions at the hadronic scale $\mu_0^2$ are assumed to be valence quarks 
and gluons. We will introduce the gluons later on in a phenomenological
way, let us therefore now restrict our discussion to the quarks. The quark
distribution is determined by the quark momentum density \cite{noi2}, 

\begin{equation} 
q_{\rm V}^{\uparrow\,(\downarrow)}(x,\mu_0^2) = {1 \over (1-x)^2}\,
\int d^3k\,\,n_q^{\uparrow\,(\downarrow)}({\bf k}^2)\,\delta
\left({x \over 1-x} - {k^+\over M_N}\right)\,\,,
\label{DVal}
\end{equation}
where
$k^+=\sqrt{{\bf k}^2+m_q^2}+k_z$ is the light-cone momentum of the struck
parton, $M_N$ and $m_q$ are the nucleon and (constituent) quark masses 
respectively and $n_q^\uparrow({\bf k}^2)$, $n_q^\downarrow({\bf k}^2)$ 
represent the density momentum distributions of the valence quark of 
$q$-flavor {\it aligned} ($\uparrow$) or {\it anti-aligned} ($\downarrow$) 
to the total spin of the parent nucleon. 
\begin{equation}
n^{\uparrow\,(\downarrow)}_{u(d)}({\bf k}^2)=\langle
N,J_z=+1/2|\sum_{i=1}^3\,
{1+(-)\tau_i^z \over 2}\,{1+(-)\sigma_i^z \over 2}\,
\delta({\bf k}-{\bf k}_i)|N,J_z=+1/2\rangle  .
\label{n_up/down}
\end{equation}

The distributions (\ref{n_up/down}) have been evaluated in the past 
making use of non-relativistic constituent quark models 
while, in the present investigation, we want to improve their
description by including relativistic effects as dictated by
a light-front formulation of a three-body interacting system.

We just outline here the basic features of this formulation 
(see \cite{lf} for a review).
In light-front dynamics the intrinsic momenta of the constituent
quarks ($k_i$) can be obtained from the corresponding momenta ($p_i$) in a generic reference 
frame through a light-front boost 
($k_i = {\cal L}_f^{-1}(P_{\rm tot})\,p_i$, $P_{\rm tot} \equiv \sum_{i=1}^3\,p_i$) 
such that the Wigner rotations reduce to identities.
The spin and spatial degrees of freedom are described by the wave
function:

\begin{equation}
\Psi=\frac{1}{\sqrt{P^+}}\,\delta(\tilde{P}-\tilde{p})\,
\chi({\bf k}_1,\mu_1,\dots,{\bf k}_3,\mu_3)
\end{equation}

\noindent
where $\mu_i$ refer to the eigenvalue of the light-front spin, 
so that the spin part
of the wave function is transformed by the tensor product of three
independent Melosh rotations: $R^{\dag}_{\rm M}({\bf k}_i, m_i)$
\cite{Melosh}, namely ${\cal R}^{\dag} = \prod_{\otimes,
i=1}^3\,R^{\dag}_{\rm M}({\bf k}_i, m_i)$.
Finally, the internal wave function is an eigenstate of the baryon mass operator
$M=M_0+V$, with  $M_0 = \sum_{i=1}^3\,\omega_i =
\sum_{i=1}^3\,\sqrt{{\bf k}_i^2+m_i^2}$,
where the interaction term $V$ must be independent on the total momentum
$\tilde{P}$ and invariant under rotations.

In the present work we will discuss results of a confining mass equation 
of the following kind
\begin{equation}
\left(M_0 + V\right)\,\psi_{0,0}(\xi) \equiv \left(\sum_{i=1}^3\,
\sqrt{{\bf k}_i^2+m_i^2} -{\tau \over \xi} + \kappa_l \,\xi\,\right)
\,\psi_{0,0}(\xi) = M\,\psi_{0,0}(\xi)\,\,,
\label{massop}
\end{equation}
where $\sum_i {\bf k}_i = 0$, $\xi = \sqrt{\vec \rho\,^2 + \vec \lambda\,^2}$ 
is the radius of the hyper-sphere in six dimension and $\vec \rho$ and 
$\vec \lambda$ are the intrinsic Jacobi coordinates 
$\vec \rho = ({\bf r}_1 - {\bf r}_2)/\sqrt2$, 
$\vec \lambda =({\bf r}_1 + {\bf r}_2 -2\,{\bf r}_3)/\sqrt6$. 
The choice of the mass operator (\ref{massop}) has been motivated by
the fact that it combines a simple form with a reasonable description
of the baryonic spectrum \cite{TBM95,CoeDaRi97}.  

The intrinsic wave function (disregarding the color part) of the
nucleon can be written: 

\begin{equation}
|N, J, J_n = +1/2 \rangle = \psi_{0,0}(\xi)\,{\cal Y}\,^{(0,0)}_{[0,0,0]}(\Omega)\,
\,\left[\chi_{MS} \phi_{MS} + \chi_{MA} \phi_{MA}\right]/\sqrt {2}
\end{equation}

\noindent where $\psi_{\gamma,\nu}(\xi)$ is the hyper-radial wave
function solution of
Eq. (\ref{massop}), ${\cal Y}\,^{(L,M)}_{[\gamma,l_\rho,l_\lambda]}(\Omega)$
the hyper-spherical harmonics defined in the hyper-sphere of unitary radius,
and $\phi$ and $\chi$ the flavor and spin wave function of mixed $SU(2)$
symmetry. Let us note that, in order to preserve relativistic covariance, 
the spin wave functions
\begin{equation}
\chi_{MS} = {1 \over \sqrt 6}\,
\left[2\,\uparrow \uparrow \downarrow - \left(\uparrow \downarrow + 
\downarrow \uparrow\right)\uparrow\right]\;;
\hspace{10mm}
\chi_{MA} = {1\over \sqrt 2}\,
\left(\uparrow \downarrow - \downarrow \uparrow\right)\uparrow
\label{spinwf} \,\,,
\end{equation}
have to be formulated by means of the appropriate Melosh transformation of the
i$th$ quark spin wave function, i.e., each individual spin vector must
be Melosh-rotated:

\begin{equation}
\chi_i = D^{1/2}[R_M(\vec{k}_i)] \chi_i^c = \frac{(m_i+\omega_i+k_{i_z})
- i \vec{\sigma}^{(i)} (\hat{z}
\times \vec{k}_{i_\perp})}{( (m_i+\omega_i+k_{i_z})^2 +
\vec{k}_{i_\perp}^{\,2})^{1/2}} \chi_i^c
\label{melosh2}
\end{equation}

\noindent where $\chi_i^c$ are the usual Pauli spinors for the particle
$i$.

The calculation of the helicity distribution at the hadronic scale
$\mu_0^2$ can be written, according to the expression
(\ref{DVal}), as:

\begin{eqnarray}
 g_1^a(x,\mu_0^2) & = & 
{\pi\over 9}{M_N \over (1-x)^2} \nonumber  \\
 & & \int_0^\infty  d\vec{k}_\perp^{\,2} \, n(\tilde
k_z^2,\vec{k}_\perp^{\,2})\frac{B_a \left(
m + M_N \frac{x}{(1-x)}\right)^2 
+ C_a \vec{k}_\perp^{\,2}}
{\left(m + M_N \frac{x}{(1-x)}\right)^2 +\vec{k}_\perp^{\,2}}
D(\tilde k_z,x)
\label{q-melosh}
\end{eqnarray}

\noindent
where 

\begin{eqnarray}
D(\tilde k_z,x) & = & \frac{M_N \frac{x}{(1-x)}-\tilde k_z}{\left|
M_N \frac{x}{(1-x)} -2 \tilde k_z\right|} \\ 
\tilde k_z(x,\vec{k}_\perp^{\,2}) & = & \frac{M_N}{2} \left[
\,\frac{(\vec{k}_\perp^{\,2}+m^2)}{M_N^2}  \frac{1-x}{x}
-\frac{x}{(1-x)} \right] 
\end{eqnarray}

\noindent 
 and  $n(\tilde k_z^2, {\bf k}^2_{\perp})$ is the total
(unpolarized and flavorless) momentum density distribution of the valence quarks in the nucleon
calculated making use of the eigenfunction $\psi_{0\,0}$ of 
Eq.~(\ref{massop}) and properly normalized to the number of particles.
The coefficients $B_a$ and $C_a$ take the values
$B_u = -C_u = 4$ and  $B_d = -C_d = -1$.

The mass equation (\ref{massop}) has been solved numerically by
fitting the parameters of the potential to reproduce the basic
features of the (non strange) baryonic spectrum up to $\approx 1500$
MeV, namely the position of the Roper resonance and the average value
of the $1^-$ states. We obtain \cite{FACCIOLI99}: 
$\tau = 3.3$ and $\kappa_l = 1.8$
fm$^{-2}$, to be compared with the corresponding
non-relativistic \cite{TBM95} fit $\tau = 4.59$ and $\kappa_l = 1.61$
fm$^{-2}$. The constituent quark masses have been chosen
$m_u = m_d = m_q = M_N/3$.

An important outcome of this relativistic treatment is the 
large amount of high momentum components in the wave function, that play an
important role in the evaluation of
transition and elastic form factors \cite{romalf} and in the large 
$x$ region in DIS
\cite{noi2}.

\begin{figure}[t]
\centerline{\psfig{file=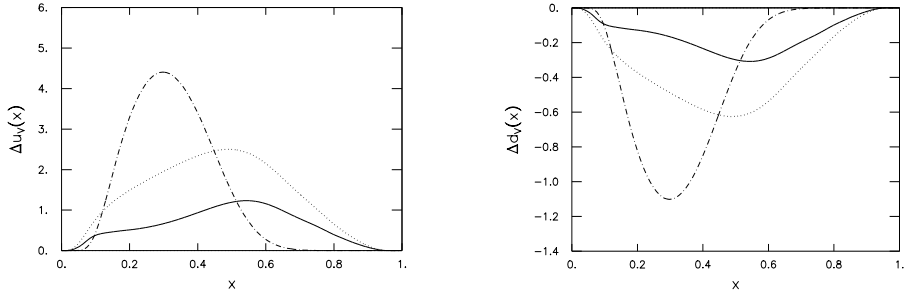,width=1\textwidth,angle=-90}}
\caption{Left panel (Fig.~1a): the polarized distribution $\Delta
u_V(x,\mu_0^2)$
as a function of $x$: the non-relativistic approximation (dot-dashed curve), and
the relativized solution of Eq.(\ref{massop})
which neglects Melosh rotations (dotted curve) are compared with the 
results of a complete light-front calculation (full curve). 
On the right panel (Fig. 1b) the distribution $\Delta d_V(x,\mu_0^2)$ (same notation).  
\label{figure1}}
\end{figure}

The relevant effects of relativistic covariance are more
evident looking at the polarized distributions
$\Delta u_V(x,\mu_0^2) \equiv g_1^u(x,\mu_0^2)$,
$\Delta d_V(x,\mu_0^2) \equiv g_1^d (x,\mu_0^2)$ 
where the spin dynamics on the light-front plays a crucial role 
(Fig. \ref{figure1}). The 
introduction of the Melosh rotations results in a substantial enhancement 
of the responses at large $x$ and in an suppression of the response for
$0.1 \stackrel{<}{\sim} x \stackrel{<}{\sim} 0.5$ as can be seen from Fig.~1. We show, in the
same figure, 
also the predictions of a pure relativized solution obtained by solving 
numerically Eq.~(\ref{massop}) and neglecting the Melosh rotation effects 
in (\ref{q-melosh}) (dotted curves). Such a calculation retains the contribution due to
the high momentum components, while the covariance requirement on the 
parton distribution is lost. 

\section{Results}

We shall give results in two scenarios characterized by different 
gluon distributions ($\Delta G$) at the hadronic scale:
\begin{itemize}
\item [i)] Scenario A : $\Delta G(x,\mu_0^2) = 0$.  
Only quark valence distributions are allowed at the hadronic scale. 
The momentum sum rule determines $\mu_0^2 = 0.094$ GeV$^2$
at NLO ([$\alpha_s(\mu_0^2)/(4\,\pi)]_{\rm NLO} = 0.142$) \cite{MaTr}.

\item [ii)] Scenario B: $\Delta G(x,\mu_0^2) = f\, G(x,\mu_0^2)$.
$f$ is the fraction of polarized gluons and has to be considered with
the appropriate sign.
\end{itemize}

A natural choice for the unpolarized gluon distribution within the present 
approach, has been discussed in refs.\cite{noi2,MaTr} and assumes a 
{\it valence-like} form \linebreak 
$G(x,\mu_0^2)= {{\cal N}_g \over 3}\,\left[ u_V(x,\mu_0^2) +
d_V(x,\mu_0^2)\right]$.
This definition implies $\int G(x,\mu_0^2)\,dx = 2$ and therefore only 
$60\%$ of the total 
momentum is carried by the valence quarks at the scale $\mu_0^2$.
In this case $\mu_0^2 = 0.220$ GeV$^2$ at NLO
\-([$\alpha_s(\mu_0^2)/(4\,\pi)]_{\rm NLO}$ \linebreak $ =0.053$).\-
If the gluons are fully polarized one has  $|\Delta G(x,\mu_0^2)| = 
G(x,\mu_0^2)$.  
Jaffe's suggestion \cite{jaffe96} in our approximation implies 
$
\Delta G(x,\mu_0^2) \approx - 0.35\,G(x,\mu_0^2)
$
\footnote{In fact in ref.\cite{jaffe96} it has
been shown that $\int \Delta G(x,\mu_0^2)\,dx < 0$. Such inequality 
does not imply $\Delta G(x,\mu_0^2) < 0$ in the whole $x$-range.}. 

\begin{figure}[t]
\centerline{\psfig{file=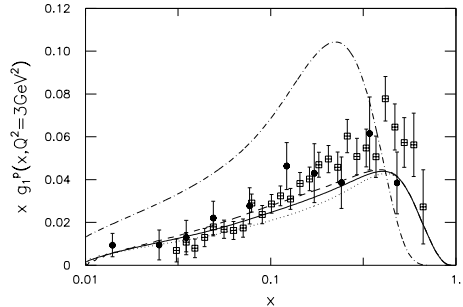,width=0.5\textwidth}}
\caption{The proton polarized structure function at $Q^2=3$ GeV$^2$.
The full curve represents the NLO ($\overline{MS}$) results of a
complete light-front calculation within a scenario where no gluons are
present at the hadronic scale (scenario A); the corresponding
non-relativistic calculation are shown by the dot-dashed line.
Scenario B is summarized by the dotted line in the case of negative
polarized gluon fraction ($\int \Delta G = -0.7$ as discussed in the
text), and by the dashed line in the case of positive gluon
polarization ($\int \Delta G = +0.7$).  Data are from SMC 
\protect\cite{SMC}, and
SLAC(E142) \protect\cite{SLACp}  experiments.
\label{figure2}}
\end{figure}

In Fig. \ref{figure2} the results for the proton structure function
$g_1^p(x,Q^2) = $ \linebreak $ \frac{1}{2} \sum_a e_a^2 (g_1^a(x,Q^2) 
+ g_1^{\bar{a}}(x,Q^2))$  are
shown and compared with the experimental data. The non-relativistic
approximation of the present calculation appears to reproduce rather poorly 
the experimental observations. Even the use of non-relativistic models
which reproduce rather well the baryon spectrum does not alter this
conclusion, as shown in ref.\cite{MaTr}. 

The full covariant calculation leads to theoretical
predictions quite close to
experimental data in the region $0.01 \stackrel{<}{\sim} x \stackrel{<}{\sim} 0.4$ under the
assumption of a pure valence component at the hadronic scale (scenario A). 
The calculation is
parameter-free and the only adjustable parameters 
($\tau$ and $\kappa_l$ in Eq.~(\ref{massop})) have been fixed to reproduce
the low-lying nucleon spectrum as discussed in the previous section.
The effect of 
relativistic covariance in the quark wave function is mainly associated to
the spin
dynamics induced by the Melosh rotations 
(\ref{q-melosh}) \cite{FACCIOLI99} and these transformations lead to a
strong suppression of this structure function in
the small-$x$ region ($x \stackrel{<}{\sim} 0.5$).

In order to introduce the gluons non perturbatively we evolve the
unpolarized distributions predicted 
by the scenario A, up to the scale of scenario B where 60\% of the total 
momentum is carried by valence quarks. At that scale we substitute the
bremsstrahlung gluons by the valence gluons as defined previously.
Moreover,  at that scale the fraction of polarized gluons  is chosen to be
{\it negative} according to Jaffe's result \cite{jaffe96}. Note that we
are maximizing the difference with respect to the radiative gluons,
because those lead to a positive polarization. Fig.~2 leads us to
conclude that the low-$x$ data on $g_1^p$ do not constrain the gluon
polarization 
strongly. If we vary the fraction of polarized gluons from 35\% to
100\% the 
quality of the agreement is  deteriorated in the region 
$0.01\stackrel{<}{\sim} x \stackrel{<}{\sim} 0.4$, but only slightly. 
For larger values of $x$
the valence 
contribution plays a major role and the behavior of the structure functions 
depends  largely on the model wave functions. 

\begin{figure}[t]
\centerline{\psfig{file=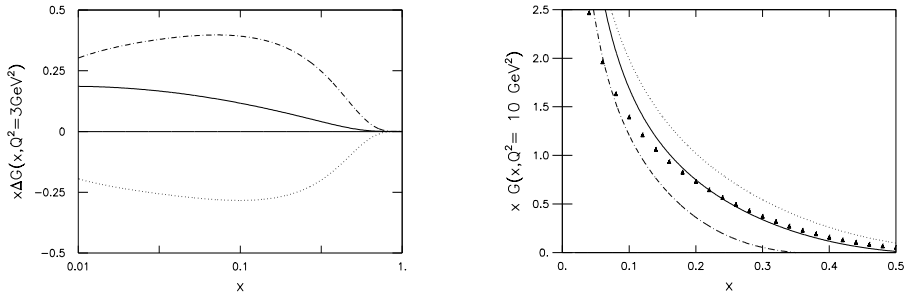,width=1\textwidth,angle=-90}}
\caption{Left panel: Polarized gluon distributions at $Q^2=3$ GeV$^2$
obtained evolving at NLO ($\overline{MS}$) the polarized partons in
both scenarios. Scenario A (full line).  Scenario B is summarized by
the dotted line in the case of negative polarized gluon fraction
($\int \Delta G(x,\mu_0^2) = -0.7$ as discussed in the text), and by
the dot-dashed line in the case of positive gluon polarization ($\int
\Delta G(x,\mu_0^2) = +0.7$). Right panel: unpolarized gluon
distribution at $Q^2=10$ GeV$^2$ obtained evolving at NLO (DIS)
unpolarized partons. Scenario A (full line); scenario B (dotted line).
For comparison also the results at LO are shown in this case
(dot-dashed).  CTEQ4 NLO (DIS) fit of ref. \protect\cite{CTEQ4}: full triangles.
\label{figure3}}
\end{figure}

Concerning the gluon distribution, our results are summarized in
Fig. \ref{figure3}. Within scenario A the 
polarized gluon distributions remain {\it positive} at the experimental 
scale as a result  of the NLO evolution.
On the contrary a large amount of {\it negative} gluon polarization
is predicted within scenario B if one assumes anti-aligned gluons at the 
hadronic scale. The distribution is largely dependent on the  polarization 
fraction. The dotted line of Fig. \ref{figure3}a  shows the consequence, 
in the deep 
inelastic regime, of the Jaffe's calculation at low energy. The absence of 
non-perturbative sea polarization results in a huge (and probably unrealistic) 
amount of negative gluon polarization to reproduce the neutron data. 
Such a large amount of gluons is, however, inconsistent with the
unpolarized gluon distributions as it is shown in Fig. \ref{figure3}b 
(dotted line again).

\section{Transversity and Relativistic Spin Effects}

	In addition to the unpolarized and helicity parton
distributions, a third parton distribution, the so called transversity
$h_1(x,Q^2)$, is required to get a complete twist-2 description of the
momentum and spin structure of the nucleon \cite{JAFFE91}. As the most remarkable
feature $h_1$ is chiral-odd and as a consequence it decouples from
electron-nucleon DIS and other easily accessible hard processes. In
doubly polarized Drell-Yan processes ($\vec{p} \vec{p} \rightarrow
l^+l^-X$) the chirality of the partons that annhilate is uncorrelated
and, in principle, $h_1$ could be measured (experiments are
already included in the research program at HERA and RHIC \cite{SAITO98}).

	We will focus on an observable defined as the ratio between
double transverse and double longitudinal asymmetry in the cross
section:

\begin{equation}
R(x_1,x_2,Q^2) \equiv \frac{A_{TT}}{A_{LL}} \frac{1}{f(\theta,\phi)} = 
\frac{\sum_a e_a^2 h_1^a(x_1,Q^2)
h_1^{\bar{a}}(x_2,Q^2)+ (x_1 \leftrightarrow x_2)} {\sum_a e_a^2
g_1^a(x_1,Q^2) g_1^{\bar{a}}(x_2,Q^2) + (x_1 \leftrightarrow x_2)}
\label{rdef}
\end{equation}

\noindent where $f(\theta,\phi)$ is a function of the angles of the
final detected lepton, $a$ indicates the flavor of the parton, $Q^2$
is the invariant mass of the produced pair and the variables $x_1$ and
$x_2$ are defined in terms of $Q^2$, the center of mass energy
$\sqrt{s}$ and the quantity $y =arcth(Q^3/Q^0)$:

\begin{eqnarray}
x_1 = \sqrt{\frac{Q^2}{s}} e^y & , & x_2 = \sqrt{\frac{Q^2}{s}} e^{-y}
\label{xdefs}
\end{eqnarray}

	The physical meaning of $h_1$ is similar to that of $g_1$ but
in a transverse basis. It is clear that in any non-relativistic
description of the nucleon, where motion and spin commutes,
$h_1^a(x,\mu_0^2) = g_1^a(x,\mu_0^2)$ at the hadronic scale
$\mu_0$. However, Melosh rotation is able to break this
degeneracy and $h_1(x,\mu_0^2)$ can be expressed in the same form as
$g_1(x,\mu_0^2)$, eq. (\ref{q-melosh}), but with different coefficients:
$B_u=4, C_u=0$ and $B_d=-1, C_d=0$ \cite{CANO99}.

\begin{figure}[t]
\begin{center}
\begin{tabular}{lr}
\psfig{file=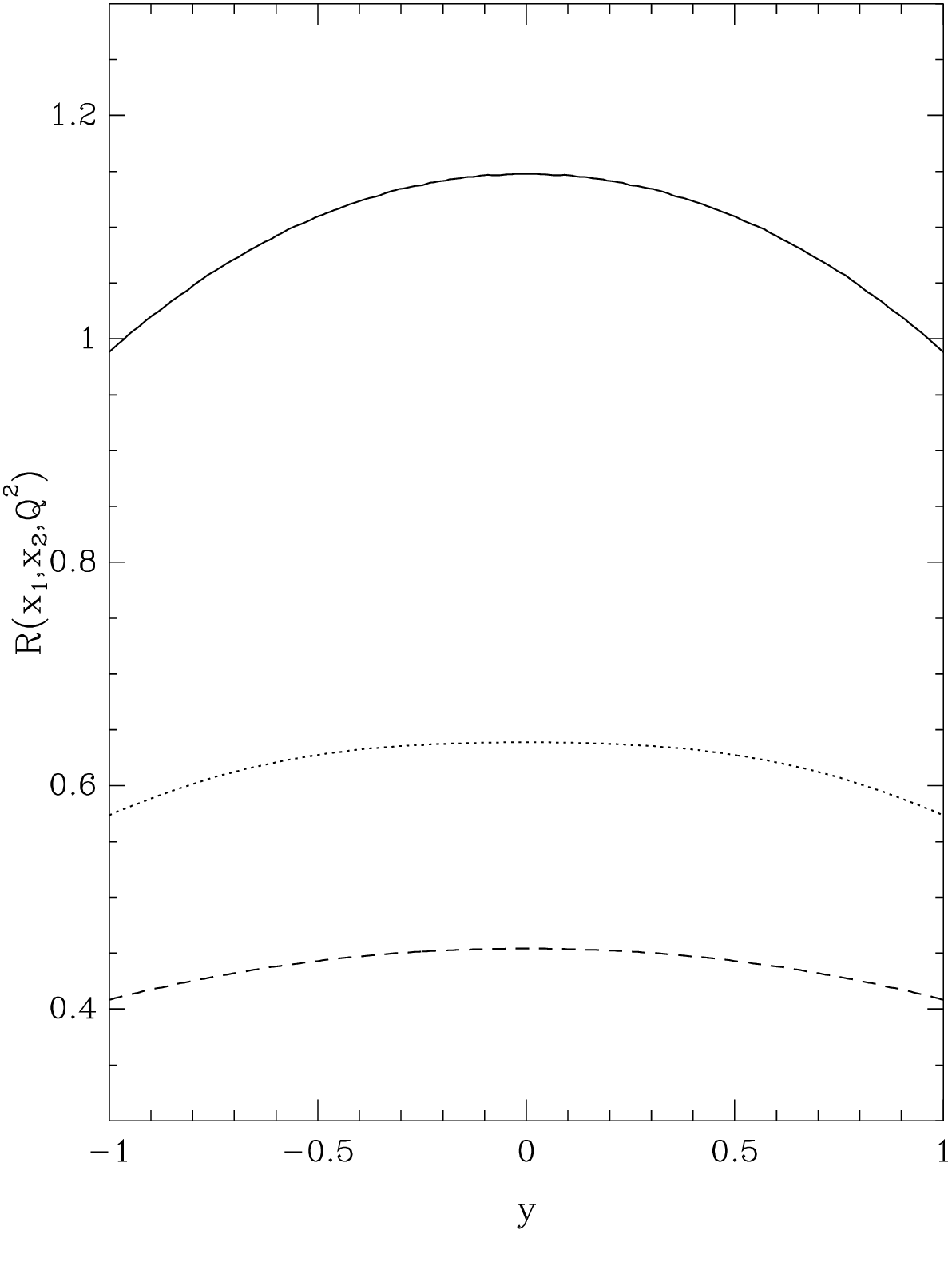,width=0.40\textwidth} & 
\psfig{file=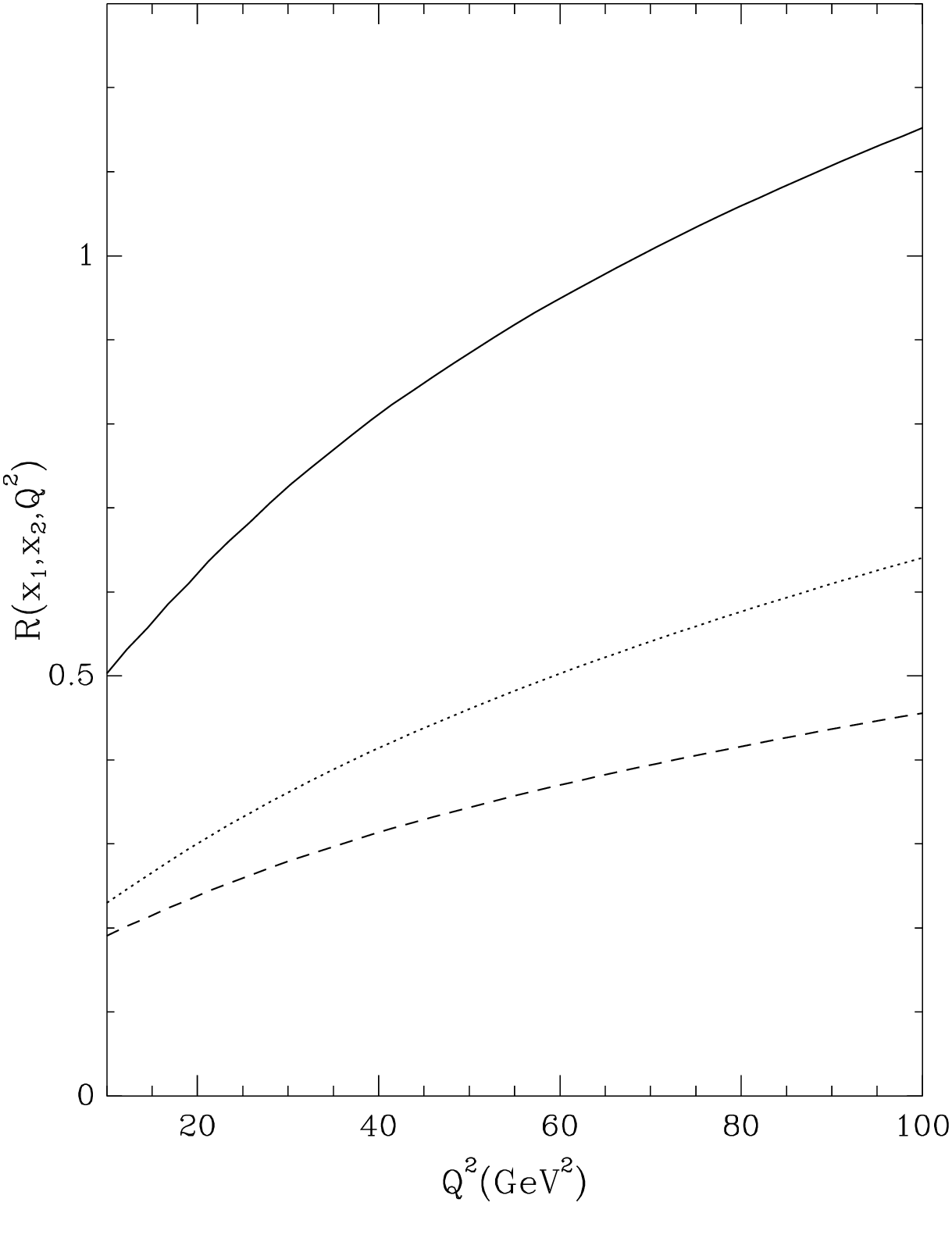,width=0.41\textwidth}
\end{tabular}
\end{center}
\caption{Ratio between transverse and longitudinal
asymmetries (eq (\ref{rdef})) as a function a) of the center of mass
rapidity $y$ for $Q^2=100$ GeV$^2$ and a center of mass energy
$\sqrt{s} = 100$ GeV (solid line); b)
of the invariant mass of the produced lepton
pair ($Q^2$) for a center of mass rapidity $y=0$ and a center of mass
energy $\sqrt{s} = 100$ GeV. The dashed line corresponds to the
case when the Melosh rotation is switched off.  The dotted line is the
result obtained in a fully non-relativistic approach.
\label{figure4}}
\end{figure}

	By employing the wave functions calculated in the previous
sections and NLO evolution we have calculated $R(x_1,x_2,Q^2)$ at a
scale $Q^2=100$ GeV$^2$ (Fig. \ref{figure4}a). Comparing the curves
shown in Fig. \ref{figure4}a it is clear that if MR is not taken into
account, i.e. we start from the equality $h_1^a(x,\mu_0^2) =
g_1^a(x,\mu_0^2)$, the obtained value for $R$ is approximately a
factor 2 smaller than when MR is considered. The same conclusion holds
for a wide kinematic regimen (see Fig.  \ref{figure4}b). Furthermore,
the chosen observable is quite insensitive to the details of the mass
operator. To check the dependence on the chosen spatial wave function
we have recalculated $R(x_1,x_2,Q^2)$ with the non-relativistic
version of the mass equation (\ref{massop}) and results are also shown in
Fig. \ref{figure4}. Therefore we can conclude that the chosen
observables is a rather direct measure of the importance of
relativistic spin effects in the nucleon (MR) while it is quite
insensitive to the details of the spatial wave function.

\section*{Acknowledgments}
We acknowledge useful conversations with S. Scopetta regarding factorization 
schemes. This work has been supported in part by DGICYT-PB94-0080
and the TMR programme of the European Commission ERB FMRX-CT96-008.

\end{document}